\documentclass[letterpaper]{article}

\usepackage{natbib,alifeconf}  
\usepackage{amsmath}
\usepackage{xcolor}
\usepackage{hyperref}
\usepackage{textcomp}

\makeatletter
\newcommand\xleftrightarrow[2][]{%
  \ext@arrow 9999{\longleftrightarrowfill@}{#1}{#2}}
\newcommand\longleftrightarrowfill@{%
  \arrowfill@\leftarrow\relbar\rightarrow}
\makeatother

\title{Towards open-ended evolutionary simulator for developing novel tumour drug delivery systems}
\author{Igor Balaz$^{1}$, Tara Petri\'c$^{1}$, \and Namid Stillman$^{2}$ \\
\mbox{}\\
$^1$Laboratory of Meteorology, Biophysics and Physics, University of Novi Sad, Serbia\\
$^2$University of Bristol, UK\\
igor.balaz@df.uns.ac.rs} 

\begin{document}
\maketitle

\begin{abstract}
Tumours behave as moving targets that can evade chemotherapeutic treatments by rapidly acquiring resistance via various mechanisms. In Balaz et al. (2021, Biosystems; 199:104290) we initiated the development of the agent-based open-ended evolutionary simulator of novel drug delivery systems (DDS). It is an agent-based simulator where evolvable agents can change their perception of the environment and thus adapt to tumour mutations. Here we mapped the parameters of evolvable agent properties to the realistic biochemical boundaries and test their efficacy by simulating their behaviour at the cell scale using the stochastic simulator, STEPS. We show that the shape of the parameter space evolved in our simulator is comparable to those obtained by the rational design.
\end{abstract}

\section{Introduction}

Precision oncology is a novel approach in tumour treatments that uses patient-specific information about the tumour to define the best possible targeted treatment \citep{Ashley2016}. Even when the right drug is identified for the right tumour profile, the problem remains of how to develop an appropriate drug delivery system that will pass all physiological barriers and reach the tumour in a high enough dose to be efficacious \citep{Stillman2020}. The sources of difficulty are numerous \citep{Richardson2020, Germain2020}. Here, we will focus only on the problem of optimal parameter identification. Our question is how to identify an optimal set of parameters, such as binding affinity, uptake, and diffusion rate in order to reach desired treatment efficacy, having in mind that tumour can rapidly develop drug resistance. With the recent increase in clinical research on developing targeted therapies with multiple drugs to deal with the adaptable tumour, two modelling problems came into focus. The first one is how to design optimal treatment that includes several drugs \citep{Tsompanas2020, Tsompanas2021}. The second, more general problem, is how to design treatment for the adaptable tumour even when future phenotypes are unknown.  Recently we have developed a simple agent-based, open-ended evolutionary engine to automatically design combinatorial oncological treatments \citep{Balaz2021}. There, we demonstrated that open-ended evolution coupled with multidimensional optimization can lead to the emergence of successful virtual therapies. Here we introduce several improvements described below.

\section{Model description and simulation results after mapping parameters to realistic values}

\begin{figure*}[t]
\begin{center}
\includegraphics[width=6.5 in]{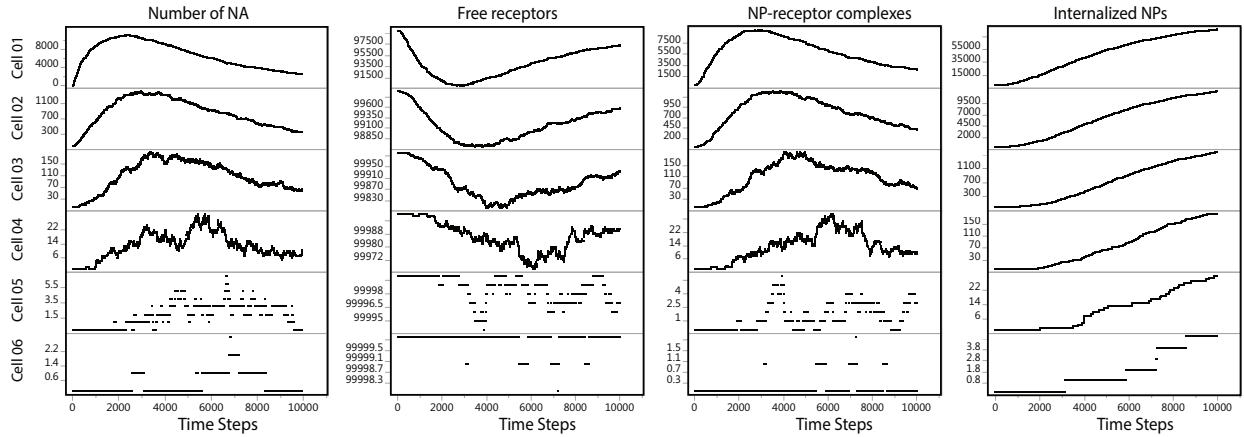}
\caption{Efficacy of NA when injected into STEPS simulator. We simulated 22-cell widths deep, counting from the vasculature point of NP extravasation. Only the first $6$ cells are shown since NPs didn't penetrate deeper. Simulated time is 11h.}
\label{fig1}
\end{center}
\end{figure*}

Details of the original model and the range of all parameters are given in \citet{Balaz2021} and here we only give a brief overview. In the model we previously introduced the simulated world is represented as a 2D grid where evolvable nano-agents (NA) learn to eliminate tumour cells (CC) while not harming healthy cells (HC).  NA can move, have internal memory, and can learn about their environment. They can only ``see'' the contents of the grid position they occupy. When the NA ``steps-on'' a cell, it will check whether it already has it in its memory (by looking at the visible properties of the cells), and memorize it if not. If the memory is full - the new cell information will take place of the oldest memorized cell. At the beginning of the simulation all agent memories are empty. Interaction of NA agents with cell-agents is modelled as a Michaelis-Menten reaction network:
  \begin{eqnarray}
\mathit{NA}_f=R \overset{p_a, p_d} {\longleftrightarrow} C \overset{p_i}{\longrightarrow} \mathit{NA}_i + R, \label{eq1}
\end{eqnarray}
where $\mathit{NA}_f$ is a free NA, $R$ is a receptor at the cell surface, $C$ is a NA-cell complex and $\mathit{NA}_i$ is an internalized NA. Probabilities of association ($p_a$), disassociation ($p_d$) and internalization ($p_i$) govern the dynamics of the NA-cell interaction. If NA encounters a ``familiar cell'', the probability of association with that agent is $p_a$.   Otherwise, the probability of association is reduced by multiplying $p_a$ with the ``curiosity'' factor.
The local fitness value is assigned to each NA agent as a sum of killed CC minus the sum of killed HC. After each $10$ time steps selection/mutation routine takes place. Mutable parameters are movement speed, $p_a$, $p_d$, $p_i$ and probability of killing ($p_k$).
CCs can evade NA by mutating their visible properties by which NA recognize them. To mimic tumour resistance, 10$\%$ of the total number of tumour cells were chosen randomly and assigned randomly-chosen resistance modifiers (one per cell). Resistance modifiers can change $p_a$, $p_d$, $p_i$, and $p_k$ of NA that interact with them. Resistance strength is randomly chosen in the interval of $30-80\%$ and is inherited after cell division.

The novel feature we introduce here is the split of the model into two modes: the learning mode and the simulation mode. In the learning mode, the goal is to find the best collective strategy to fight the tumour, regardless of the duration of the treatment. After interacting with a cell, NA agents do not perish but continue with movement. We ran the learning mode for $10,000$ time steps and the efficacy of top performing NAs is evaluated in the ``simulation mode'' and in the STEPS simulator.
In the \textit{simulation mode} we include the realistic temporal dimension based on the following assumptions: (i) one grid cell equals one cell agent, (ii) average tumour cell diameter is 10 {\textmu}m, (iii) the diffusion coefficient range for nanoparticles in a fluid is in the range of $10^{-10} cm^{2}\slash s$ \citep{Katayama2009, Pinheiro2007}. Given that we are in 2D space, NA then needs $t = (10^{-3})^{2} \div 2 \times 10^{-10} = 0.5 \times 10^{4}$ seconds to move from one empty grid cell to a neighbouring cell. As we fixed NA speed in an empty cell to $1$, one time-step in the simulation mode equals $5000$ seconds.
To calculate injection dynamics of NAs in \textit{simulation mode}, we took pharmacokinetics values for a typical anticancer drug Doxorubicin \citep{Dawidczyk2017}. There, a drug reaches maximum concentration in the tumour in 20 hours ($\sim$ 14 simulated time steps) after the injection. For the following 80 hours ($\sim$ 72 time steps) a drug concentration slowly declines until it reaches zero. Therefore, during the first 14 simulation steps, at each step we inject 1/14 of the predefined total dose. As expected, most efficacious option is when the maximum tolerated dose of NA is injected. This eliminated $7\%$ of tumour cells after the single dose treatment (data not shown).
To test how the spatial distribution of cells affects nanoparticle transport, we used the Stochastic Engine for Pathway Simulation (STEPS) \citep{Wils2009} where we incorporates spatiality through the discretisation of the domain into tetrahedral well-mixed subregions. To normalize evolved NA parameters to realistic values we take the range of association rate constant ($ka$) to be between $10^4$ and $10^6$ [$1$/Ms] \citep{Hauert2013}. Since the time step is $5,000$ seconds, $ka$ is $5x10^7$ to $5x10^9$ [1/M time step]. If we assume that one NA represents $10^5$ particles ($1.66x10^-7$ [M]), then the $ka$ range is $8.3$ to $8.3x10^2$ [particles / time step]. The normalization factor is then $1.66x10^-7$ [M] / $5,000$ [$s$] = $1.2x10^3$ . Therefore, for example $p_a = 0.3$ translates to ka=$3.61x10^2$ [$1$/Ms]. Using the same approach we calculated the normalization factor for disassociation constant $kd$ and internalization constant ($ki$) to be $2x10^{-4}$. In the simulations, maximum penetration depth is $6$ cells Fig.\ref{fig1} and NA were able to kill all cells into which they were internalized.

\section{Conclusions}
At this stage it is hard to directly compare our results to clinically relevant ones due to the differences in tumour size scaling, drug-action mechanisms and influence of tumour heterogeneity on drug diffusion. Nevertheless, it is important to note that the distribution of NA parameters we obtained as a result of \textit{in silico} evolution is similar to those used to synthesize nanoparticle-based DDS (which is result of highly skilled rational design) as well as to parameters obtained via deterministic modelling \citep{Hauert2013}. Therefore we believe that our evolutionary approach can generate useful leads in designing novel drug-delivery systems, especially since our engine is designed to deal with tumour changeability which is one of the main problems in long term efficacy of oncology treatments. Therefore, our ongoing work is focused on further closing the gap between our evolvable simulator and pre-clinical investigations, mostly by developing closer mapping between the parameter space of the simulator and the corresponding biochemical counterparts.

\section{Acknowledgements}

This project has received funding from the European Union’s Horizon 2020 research and innovation program under grant agreement No 800983.

\footnotesize
\bibliographystyle{apalike}
\bibliography{example} 

\begin{thebibliography}{}

\bibitem[Ashely, 2016]{Ashley2016}
Ashely, E. (2016).
\newblock Towards precision medicine.
\newblock {\em Nat Rev Genet}, 17:507–522.

\bibitem[Balaz et~al., 2021]{Balaz2021}
Balaz, I., Petri{\'c}, T., Kovacevic, M., Tsompanas, M.-A., and Stillman, N.
  (2021).
\newblock Harnessing adaptive novelty for automated generation of cancer
  treatments.
\newblock {\em Biosystems}, 199:104290.

\bibitem[Dawidczyk et~al., 2017]{Dawidczyk2017}
Dawidczyk, C., Russell, L., Hultz, M., and Searson, P. (2017).
\newblock Tumor accumulation of liposomal doxorubicin in three murine models:
  optimizing delivery efficiency.
\newblock {\em Nanomedicine}, 13:1637–1644.

\bibitem[Germain et~al., 2020]{Germain2020}
Germain, M., Caputo, F., Metcalfe, S., Tosi, G., Spring, K., Aslund, A.,
  Pottier, A., Schiffelers, R., Ceccaldi, A., and Schmid, R. (2020).
\newblock Delivering the power of nanomedicine to patients today.
\newblock {\em Journal of Controlled Release}, 326:164--171.

\bibitem[Hauert et~al., 2013]{Hauert2013}
Hauert, S., Berman, S., Nagpal, R., and Bhatia, S. (2013).
\newblock A computational framework for identifying design guidelines to
  increase the penetration of targeted nanoparticles into tumors.
\newblock {\em Nano today}, 8:566–576.

\bibitem[Katayama et~al., 2009]{Katayama2009}
Katayama, K., Nomura, H., Ogata, H., and Eitoku, T. (2009).
\newblock Diffusion coefficients for nanoparticles under flow and stop-flow
  conditions.
\newblock {\em Physical Chemistry Chemical Physics}, 11:10494–10499.

\bibitem[Pinheiro et~al., 2007]{Pinheiro2007}
Pinheiro, J.~P., Domingos, R., Lopez, R., Brayner, R., Fi{\'e}vet, F., and
  Wilkinson, K. (2007).
\newblock Determination of diffusion coefficients of nanoparticles and humic
  substances using scanning stripping chronopotentiometry (sscp).
\newblock {\em Colloids and Surfaces A: Physicochemical and Engineering
  Aspects}, 295(1-3):200--208.

\bibitem[Richardson and Caruso, 2020]{Richardson2020}
Richardson, J. and Caruso, F. (2020).
\newblock Nanomedicine toward 2040.
\newblock {\em Nano Lett}, 20:1481–1482.

\bibitem[Stillman et~al., 2020]{Stillman2020}
Stillman, N., Kovacevic, M., Balaz, I., and Hauert, S. (2020).
\newblock In silico modelling of cancer nanomedicine, across scales and
  transport barriers.
\newblock {\em npj Comput Mater}, 6:92.

\bibitem[Tsompanas et~al., 2020]{Tsompanas2020}
Tsompanas, M.-A., Bull, L., Adamatzky, A., and Balaz, I. (2020).
\newblock Novelty search employed into the development of cancer treatment
  simulations.
\newblock {\em Informatics in Medicine Unlocked}, 19:100347.

\bibitem[Tsompanas et~al., 2021]{Tsompanas2021}
Tsompanas, M.-A., Bull, L., Adamatzky, A., and Balaz, I. (2021).
\newblock In silico optimization of cancer therapies with multiple types of
  nanoparticles applied at diffeerent times.
\newblock {\em Computer Methods and Programs in Biomedicine}, 200:105886.

\bibitem[Wils and De~Schutter, 2009]{Wils2009}
Wils, S. and De~Schutter, E. (2009).
\newblock Steps: Modeling and simulating complex reaction-diffusion systems
  with python.
\newblock {\em Frontiers in Neuroinformatics}, 3:15.

\end{thebibliography}

\end{document}